\documentclass{PoS}

\PoS{PoS(WC2004)}

\title{Some Mathematical Aspects of the Lifshitz Formula for the Thermal
Casimir
Force}

\ShortTitle{Mathematical Aspects of the Lifshitz Formula}

\author{\speaker{V.~M.~Mostepanenko}\thanks{On leave from Noncommercial
Partnership ''Scientific Instruments'', Moscow, Russia.}\,\,, A.~O.~Caride,  
G.~L.~Klimchitskaya\thanks{On leave from North-West Technical University,
St.Petersburg, Russia.}{\ } and S.~I.~Zanette \\
        Centro Brasileiro de Pesquisas F\'{\i}sicas,
 Rio de Janeiro, Brazil\\
        E-mail: \email{Vladimir.Mostepanenko@itp.uni-leipzig.de}, 
\email{caride@cbpf.br}, 
\email{Galina.Klimchitskaya@itp.uni-leipzig.de}, \email{szanette@cbpf.br}}

\abstract{The applicability of the Lifshitz formula is discussed
to the case of two thick parallel plates made of real metal.
The usual description of the zero-point vacuum oscillations 
on the background of the frequency-dependent dielectric
permittivity is shown to be in contradiction with thermodynamics.
Instead, the Lifshitz formula for the Casimir free energy should 
be reformulated in terms of the reflection coefficients 
containing the surface impedance instead of the dielectric
permittivity. This approach is presently confirmed experimentally
by precision measurements of the van der Waals and Casimir forces
in micromechanical systems and it is in agreement with thermodynamics. 
}

\FullConference{Fourth International Winter Conference on Mathematical Methods
in Physics\\
                 09 - 13 August 2004\\
                 Centro Brasileiro de Pesquisas Fisicas (CBPF/MCT), Rio de
Janeiro, Brazil}

\begin{document}
\def \beq {\begin{equation}}
\def \eeq {\end{equation}}
\def \bes {\begin{eqnarray}}
\def \ees {\end{eqnarray}}
\def \ni {\noindent}
\def \nn {\nonumber}
\def \z {\tilde{z}}
\def \kp {k_{\bot}}
\def \e {\varepsilon}
\def \dpa {\Delta_{\|}}
\def \dpp {\Delta_{\bot}}

\section{Introduction}

It is common knowledge that Lifshitz formula \cite{19}
describes the van der Waals and Casimir force acting between
two thick plane parallel material plates separated by a gap
of width $a$. According to this formula, the free energy
of the van der Waals and Casimir interaction can be
represented in terms of reflection coefficients
    \beq
{\cal{F}_R}=\frac{k_BT}{2\pi}
\int_0^{\infty}\!\!\!{k_{\!\bot}\,dk_{\!\bot}}
\sum\limits_{l=0}^{\infty}{\vphantom{\sum}}^{\prime}\left\{
\ln\left[1-r_{\|}^{2}(\xi_l,k_{\!\bot})e^{-2aq_l}\right] 
+
\ln\left[
\vphantom{r_{\|}^{2}(\xi_l,k_{\!\bot})e^{-2aq_l}}
1-r_{\bot}^{2}(\xi_l,k_{\!\bot})e^{-2aq_l}\right]
\right\}.
\label{e1}
\eeq
\ni
Here prime means the addition of a multiple 1/2 near the term with
$l=0$, and
the Lifshitz reflection
coefficients take the form
    \bes
&& r_{\|}^{2}(\xi_l,k_{\!\bot})\equiv
r_{\|,L}^{2}(\xi_l,k_{\!\bot})=
\left(\frac{\varepsilon_lq_l-k_l}{\varepsilon_lq_l+k_l}\right)^2,
\nn \\
&& r_{\bot}^{2}(\xi_l,k_{\!\bot})\equiv
r_{\bot,L}^{2}(\xi_l,k_{\!\bot})=
\left(\frac{q_l-k_l}{q_l+k_l}\right)^2,
\label{e2}
\ees \ni
where
$\varepsilon_l\equiv\varepsilon(i\xi_l)$, $\varepsilon(\omega)$ is
the dielectric permittivity of the plate material, 
$\xi_l=2\pi k_B Tl/\hbar$ ($l=0,1,2,\ldots$) are the Matsubara
frequencies, and
$k_l^2\equiv k_{\!\bot}^2+\varepsilon_l\xi_l^2/c^2$,
{\ }$q_l^2\equiv k_{\!\bot}^2+\xi_l^2/c^2$.

Beginning in 2000, the behavior of the thermal correction to the
Casimir force between real metals has been hotly debated. It was shown
that  Lifshitz formula
leads to different results depending on the model of metal
conductivity used. For real metals at low frequencies $\omega$,
the dielectric permittivity $\varepsilon$ varies as $\omega^{-1}$. After
substituting $\varepsilon\sim\omega^{-1}$
 into the Lifshitz formula, the result is a thermal
correction which is several hundred times greater than for
ideal metals at separations of a few tenths of a micrometer \cite{3,8}
The attempt \cite{6} to modify the zero-frequency term of the Lifshitz
formula for real metals, assuming that it behaves as in the case of
ideal metals, also leads
to a large thermal correction to the Casimir force at short separations.

It is important to note that in the
approaches of both  \cite{3,8} and also of
 \cite{6} a thermodymanic puzzle
arises, i.e., the Nernst heat theorem is violated for a perfect
lattice \cite{12,17}. (See also  \cite{16} where it is shown that for
the preservation of the Nernst heat theorem in the approach of
 \cite{3,8} it is necessary to have metals with
defects or impurities; it is common knowledge, however, that thermodynamics
must be valid for both perfect and imperfect lattices.)
This puzzle casts doubt on the many applications of the Lifshitz theory
of dispersion forces, and thus represents a potentially serious challenge to
both experimental and theoretical physics. By contrast, the use of
$\varepsilon\sim\omega^{-2}$, as holds in a free electron plasma
model neglecting relaxation, leads \cite{5,4} to a small thermal
correction to the Casimir force at short separations.
This is in qualitative agreement with
the case of an ideal metal and is consistent with the Nernst heat
theorem. It should be borne in mind, however, that the plasma model
is not universal, and is applicable only in the case when the characteristic
frequency is in the domain of infrared optics.

The present paper demonstrates that the main
reason why the Drude model in combination with the Lifshitz theory
had failed to describe the thermal Casimir force is the 
inadequacy of the standard concept of a fluctuating electromagnetic 
field on the background of $\varepsilon$ depending only on
frequency inside a lossy real metal. To avoid a contradiction with
thermodynamics, one should use the reflection coefficients expressed
in terms of the surface impedance. 

\section{The fluctuating field and
the surface impedance}

The concept of a fluctuating electromagnetic
field works well for the description of zero-point oscillations
in media with a frequency-dependent dielectric permittivity
where no real electric current does arise. We will now consider a
conductor in an external electric 
field, which varies with some
frequency $\omega$ satisfying the conditions
    \beq
    l\ll\delta_n(\omega),\qquad l\ll\frac{v_F}{\omega}, \label{e3}
    \eeq
\ni where $l$ is the mean free path of a conduction electron,
$\delta_n(\omega)=c/\sqrt{2\pi\sigma\omega}$ is the penetration
depth of the field inside a metal, $\sigma$ is the conductivity,
and $v_F$ is the Fermi velocity. Eqs.~(\ref{e3}) determine the
domain of the normal skin effect \cite{21}. In this frequency
region the external field leads to the initiation of a real
current of the conduction electrons and the dielectric permittivity
is modelled by the Drude function $\varepsilon\sim\omega^{-1}$
leading to the difficulties with the Lifshitz formula mentioned
in Introduction.

The physical reason for these difficulties becomes clear
when one observes that the alternating electric field with
frequencies characteristic for the normal skin effect inevitably
leads to heating of a metal when it penetrates through the skin
layer. By contrast, the thermal photons in thermal equilibrium
with a metal plate or the virtual photons (giving rise
to the Casimir effect) can not lead to
the initiation of a real current and heating of the metal 
(this is prohibited by thermodynamics). Hence the
concept of a fluctuating electromagnetic field penetrating inside
a metal cannot describe virtual and thermal photons in the
frequency region (\ref{e3}). As a consequence, the Lifshitz formula
can not be applied in combination with the Drude dielectric
function in the domain of the normal skin effect.

As is evident from the foregoing, another theoretical basis is needed
to find the thermal Casimir force between real metals different
from the approach used in the case of dielectrics. Here we show that
this basis is given by the surface impedance boundary conditions
introduced by M.~A.~Leontovich \cite{19,24}.
The fundamental difference of the surface impedance 
boundary conditions  from
the other approaches is that they permit not to consider the
electromagnetic fluctuations inside a metal. Instead, the
following boundary conditions are imposed taking into account
the properties of real metal 
\beq
{{\mathbf{E}}}_t=Z(\omega)
\left[{{\mathbf{B}}}_t\times{{\mathbf{n}}}\right],
\label{e9} 
\eeq \ni 
where $Z(\omega)=1/\sqrt{\varepsilon(\omega)}$ is 
the Leontovich surface impedance of
the conductor, ${\mathbf{E}}_t$ and ${\mathbf{B}}_t$
are the tangential components of electric and magnetic fields, and
${\mathbf{n}}$ is the unit normal vector to the surface
(pointed inside a metal). The boundary condition (\ref{e9}) can
be used to determine the electromagnetic field outside a metal.
Note, that the impedance $Z(\omega)$ and the condition (\ref{e9})
suggest a more universal description than the one by means of
$\e$. They still hold in the domain of the anomalous skin
effect where a description in terms of the dielectric permittivity 
$\e(\omega)$ is impossible. For ideal metals it holds $Z\equiv 0$.

The use of the Leontovich impedance in Eq.~(\ref{e9}) which does not depend
on the polarization state and transverse momentum, is of prime importance.
Note that in  \cite{12n} the exact impedances
depending on a transverse momentum were used. This has led to the
same conclusions as were obtained previously from the Lifshitz formula
combined with the dielectric permittivity $\varepsilon\sim\omega^{-1}$.
As was already mentioned above, these conclusions are in violation
of the Nernst heat theorem for a perfect
lattice \cite{12,17,16}.
Although a recent review \cite{12n} claims agreement
with the Nernst heat theorem
in  \cite{3,8},  no specific objections
against the rigorous analytical proof of the opposite statement in
 \cite{17} are presented.
The fallacy in the calculations of  \cite{12n}
concerning the type of the impedance
is that it disregards the requirement that the reflection properties
for virtual photons on a classical boundary should be the same as
for real photons.  Paper \cite{17} demonstrates
in detail that by enforcing this requirement the exact and Leontovich
impedances coincide at zero frequency and lead to the conclusions
of  \cite{15} which are in perfect agreement with the Nernst
heat theorem.

\section{Lifshitz formula in terms of surface impedance}

Let us consider the case of real eigenfrequencies
$\omega_{k_{\bot},n}^{\|}$, $\omega_{k_{\bot},n}^{\bot}$ (i.e.,
the pure imaginary im\-pe\-dan\-ce). The total free energy
of the electromagnetic oscillations is given by the sum of the
free energies of oscillators over all possible values of
their quantum numbers, 
\beq {\cal{F}}= \sum\limits_{\alpha}\left[
\frac{\hbar\omega_{\alpha}}{2}+k_BT\ln
\left(1-e^{-\frac{\hbar\omega_{\alpha}}{k_BT}}\right)\right]=
k_BT \sum\limits_{\alpha}
\ln\left(2\sinh{\frac{\hbar\omega_{\alpha}}{2k_BT}}\right).
\label{e14} 
\eeq \ni 
At $T\to 0$, the value of
$\cal{F}$ from Eq.~(\ref{e14}) coincides with the sum
of the zero-point energies which is usually considered
at zero temperature.

Applying this to the electromagnetic oscillations between metal
plates, where $\alpha=\{p,{\mbox{$k$}}_{\!\bot},n\}$,
and $p=\bot$ or $\|$ labels the polarization states,
we obtain
\beq
 {\cal{F}}=k_BT
\int_0^{\infty}\frac{k_{\!\bot}\,dk_{\!\bot}}{2\pi}
\sum\limits_{n}\left[
\ln\left(2\sinh{\frac{\hbar\omega_{k_{\bot},n}^{\|}}{2k_BT}}\right)
+\ln\left(2\sinh{\frac{\hbar\omega_{k_{\bot},n}^{\bot}}{2k_BT}}\right)
    \right]. \label{e15} 
\eeq \ni
Using the impedance boundary conditions, it can be easily shown that
the eigenfrequencies of
the electromagnetic field between plates with parallel and
perpendicular polarizations are determined by the equations
\beq
\Delta_{\|}(\omega,k_{\!\bot})\equiv
\frac{1}{2}e^{-aq}\left(1-\eta^2\right)\left(\sinh aq-
\frac{2i\eta}{1-\eta^2}\cosh aq\right)=0,
\label{e16}
\eeq
\ni
\beq
\Delta_{\bot}(\omega,k_{\!\bot})\equiv
\frac{1}{2}e^{-aq}\left(1-\kappa^2\right)\left(\sinh aq+
\frac{2i\kappa}{1-\kappa^2}\cosh aq\right)=0,
\label{e17}
\eeq
\ni
where $\eta=\eta(\omega)=Z\omega/(cq)$,
$\kappa=\kappa(\omega)=Zcq/\omega$,
and $q^2=k_{\bot}^2-\omega^2/c^2$.

The expression in the right-hand side of Eq.~(\ref{e15}) is
evidently divergent. Before performing a renormalization, let us
equivalently represent the sum over the eigenfrequencies
$\omega_{k_{\bot},n}^{\|,\bot}$ by the use of the argument theorem
\cite{2}.
Then Eq.~(\ref{e15}) can be rewritten as
\beq
{\cal{F}}=k_BT
\int_0^{\infty}\frac{k_{\!\bot}\,dk_{\!\bot}}{2\pi} \frac{1}{2\pi
i}\oint_{C_1} \ln\left(2\sinh{\frac{\hbar\omega}{2k_BT}}\right)
d\left[\ln\Delta_{\|}(\omega,k_{\!\bot})+
\ln\Delta_{\bot}(\omega,k_{\!\bot})\right].
\label{e18}
\eeq
    \ni
Here, the closed contour $C_1$ is bypassed counterclockwise. It
consists of two arcs, one having an infinitely small radius
$\varepsilon$ and the other one an infinitely large radius $R$,
and two straight lines $L_1,\,L_2$ inclined at the angles $\pm 45$
degrees to the real axis. The quantities
$\Delta_{\|,\bot}(\omega,k_{\!\bot})$ have their roots at the
photon eigenfrequencies and are defined in Eqs.~(\ref{e16}) and
(\ref{e17}). Unlike the usual derivation of the Lifshitz
formula at $T\neq 0$ \cite{15n} the function under the
integral in (\ref{e18}) has branch points rather than poles at the
imaginary frequencies $\omega_l=i\xi_l$.
The contour $C_1$ is
chosen so as to avoid all these branch points and enclose all the
photon eigenfrequencies.

After the integration by parts and some rearrangement \cite{15},
we find the equivalent but more simple
expression for the Casimir free energy
    \beq
{\cal{F}}=\frac{k_BT}{2\pi}
\int_0^{\infty}{k_{\!\bot}\,dk_{\!\bot}}
\sum\limits_{l=0}^{\infty}{\vphantom{\sum}}^{\prime}\left[
\ln\Delta_{\|}(\xi_l,k_{\!\bot})+
\ln\Delta_{\bot}(\xi_l,k_{\!\bot})\right].
\label{e19}
\eeq

Expression (\ref{e19}) is still infinite. To remove the
divergences, we subtract from the right-hand side of
Eq.~(\ref{e19}) the free energy in the case of infinitely separated
interacting bodies ($a\to\infty$). Then the physical,
renormalized, free energy vanishes for infinitely remote plates.
From Eqs.~(\ref{e16}) and (\ref{e17}) after the substitution $\omega\to
i\xi_l$ in the limit $a\to\infty$ it follows
    \bes &&
\Delta_{\|}^{\!\infty}(\xi_l,k_{\!\bot})=\frac{1}{4}\left(1+\eta_l^2\right)
\left(1+\frac{2\eta_l}{1+\eta_l^2}\right),
\nn \\
&&
\Delta_{\bot}^{\!\infty}(\xi_l,k_{\!\bot})=\frac{1}{4}
\left(1+\kappa_l^2\right)
\left(1+\frac{2\kappa_l}{1+\kappa_l^2}\right).
\label{e20}
\ees
    \ni
The renormalization prescription is equivalent to the change of
$\Delta_{\|,\bot}(\xi_l,k_{\!\bot})$ in Eq.~(\ref{e19}) for
    \beq
\Delta_{\|,\bot}^{\! R}(\xi_l,k_{\!\bot})\equiv
\frac{\Delta_{\|,\bot}(\xi_l,k_{\!\bot})}{\Delta_{\|,\bot}^{\!\infty}
(\xi_l,k_{\!\bot})}=
1-r_{\|,\bot}^{2}(\xi_l,k_{\!\bot})e^{-2aq_l},
\label{e21}
\eeq
    \ni
where the quantities $r_{\|,\bot}(\xi_l,k_{\!\bot})$ have the
meaning of reflection coefficients and are given by
    \beq
r_{\|}^{2}(\xi_l,k_{\!\bot})=
\left(\frac{cq_l-Z_l\xi_l}{cq_l+Z_l\xi_l}\right)^2,
\quad
r_{\bot}^{2}(\xi_l,k_{\!\bot})=
\left(\frac{\xi_l-Z_lcq_l}{\xi_l+Z_lcq_l}\right)^2.
\label{e22}
\eeq
    \ni
Here $Z_l\equiv Z(i\xi_l)$.
The reflection coefficients (\ref{e22}) are in accordance with
 \cite{24} where the reflection of a plane electromagnetic
wave incident from vacuum onto the plane surface of a metal was
described in terms of the Leontovich surface impedance.

Thus, the final renormalized expression for the Casimir
free energy in the surface impedance approach is given once more
by the Lifshitz formula (\ref{e1}),
where the reflection coefficients are expressed in terms of
impedance according Eq.~(\ref{e22}).

The above derivation was performed under the assumption that the
photon eigenfrequencies are real. This is, however, not the case
for arbitrary complex impedance. If the photon eigenfrequencies
are complex, the free energy is not given by Eq.~(\ref{e14}) (which
is already clear from the complexity of the right-hand side of
this equation). For arbitrary complex impedance the correct
expression for the free energy should be determined from the
solution of an auxiliary electrodynamic problem \cite{31}. 
In fact the Casimir free energy is the
functional of the impedance even when the impedance has a nonzero
real part taking absorption into account. The solution of the
auxiliary electrodynamic problem leads to conclusion \cite{31}
that the correct free energy is obtained from
Eqs.~(\ref{e1}), (\ref{e22}) by analytic continuation to arbitrary
complex impedances, i.e., to arbitrary oscillation spectra. The
qualitative reason for the validity of this statement is that the
free energy depends only on the behavior of $Z(\omega)$ at the
imaginary frequency axis where $Z(\omega)$ is always real.

It must be emphasized that the
surface impedance approach is in perfect agreement with
thermodynamics. In the impedance approach the entropy, defined as
    \beq
S(a,T)=-\frac{\partial{\cal{F}}_R(a,T)}{\partial T}, 
\label{e24}
\eeq
    \ni
is positive and equal to zero at zero temperature in accordance
with the Nernst heat theorem (remind that this is not the case in
the approaches based on the use of a frequency dependent
dielectric permittivity). 

\section{Conclusions and discussion}

In the above we demonstrated that the Lifshitz formula for
the Casimir free energy can be derived starting from the boundary
condition on the surface of real metal containing the Leontovich
impedance. In doing so there is no need to use the concept of
a fluctuating electromagnetic field inside a metal. We argued
that the standard concept of a fluctuating field inside a metal, 
described by the dielectric permittivity depending only on
frequency cannot serve as an adequate model for the
zero-point oscillations and thermal photons. 
It follows from the
fact that the vacuum oscillations and thermal photons in
equilibrium can not lead to a heating of a
metal as do the electromagnetic fluctuations on the background
of $\varepsilon(\omega)$. 
If this fact is overlooked, contradictions with the
thermodynamics arise when one substitutes into the Lifshitz
formula the Drude dielectric
function taking into account the volume relaxation and,
consequently, the Joule heating.

In the impedance approach the
entropy is in all cases nonnegative and takes zero value at zero
temperature. Thus, Eqs.~(\ref{e1}) and (\ref{e22}) lay down
the theoretical foundation for the calculation of the thermal Casimir
effect. In fact the approaches of  \cite{3,8,6} and the
impedance approach of  \cite{17,15} predict quite different
magnitudes of the thermal corrections to the Casimir force.
Up to separation distances of a few hundred nanometers, the thermal
correction predicted by the impedance approach is negligibly small.
As to the thermal corrections of  \cite{3,8,6}, they may 
achieve several percent of force magnitude. Recent experiment
\cite{32} on measuring the Casimir force by means of a 
micromechanical torsional oscillator is consistent with the
theoretical predictions of the impedance approach. At the same
time, the experimental data of this experiment is in drastic
contradiction with the approaches of  \cite{3,8,6}.
The experiments on measuring the Casimir and van der Waals
forces by means of an atomic force microscope also suggest
good opportunity to distinguish between the two approaches.

To conclude, different derivations of the Lifshitz formula
for the Casimir free energy in the case of real metals
lead to one and the same mathematical expression containing,
however, two different pairs of reflection coefficients.
Recent results demonstrate quite clear that for real metals
the use of reflection coefficients, expressed in terms of the 
frequency-dependent dielectric permittivity, is not only
in violation of thermodynamics but is also in contradiction
with experiment. By this reason the reflection coefficients in
terms of the Leontovich impedance are evidently preferable. 

\section*{Acknowledgments}

V.M.M.\ is grateful to the participants of the Seminar at Centro Brasileiro
de Pesquisas F\'{\i}sicas (CCP) for discussion.
The authors acknowledge FAPERJ for financial support.

\end{document}